\documentclass[fleqn,10pt]{wlscirep}
\usepackage{amsmath}
\usepackage{amssymb}
\usepackage{graphicx}
\usepackage{dcolumn}
\usepackage{bm}
\usepackage[final]{changes} 

\usepackage[utf8]{inputenc}
\usepackage[T1]{fontenc}
\usepackage{mathptmx}
\usepackage{hyperref}
\hypersetup{%
	pdftitle={Coil of the highest possible Q},%
	pdfauthor={A. Rikhter and M. M. Fogler},%
	pdfpagemode={UseNone},%
	pdfstartview={FitH},%
	breaklinks=true,%
	citecolor=blue,%
	colorlinks=true,%
	linkcolor=blue,%
	urlcolor=blue}

\DeclareMathOperator{\im}{Im}
\DeclareMathOperator{\re}{Re}

\begin{document}
\newcommand\UCSD{Department of Physics, University of California San Diego, 9500 Gilman Drive, La Jolla, California 92093, USA}
\title{\replaced{Inductor coil of the highest possible $\mathbf{Q}$}{What is the coil of the highest possible Q?}}
\author[1, +, *]{A. Rikhter}
\author[1, +]{M. M. Fogler}
\affil[1]{\UCSD}
\affil[*]{arikhter@ucsd.edu.}
	
\date{\today}

\begin{abstract}

\textbf{Abstract}\newline 
The geometry of an inductor made of a long thin wire and having the highest possible Q-factor is found by numerical optimization.
As frequency increases, the Q-factor first grows linearly and then according to a square-root law,
while the cross-section of the optimal coil evolves from near-circular to sickle-shaped.

\end{abstract}

\maketitle
\section*{Introduction}
Given a piece of wire, how can one wind it into a coil of the maximum possible $Q$-factor?
While previously this question has been treated almost exclusively in the context of radio engineering,~\cite{Medhurst1947a, Medhurst1947b}
in this work we address it as a problem in mathematical physics.
To constrain the size of the coil, we have the following geometric parameters fixed:
the total wire length $W$, the conducting core diameter $d_i$, 
and the effective outer diameter $d$. 
We define $d$ in terms of the maximum possible wire density
$n_2 \equiv (\pi d^2 / 4)^{-1}$ per unit area.
Thus, for the hexagonal closed packing of round wires,
$d$ is $(12 / \pi^2)^{1 / 4} = 1.050$ times the actual outer diameter. 
The current is taken to be $I = e^{-i\omega t}$. 
We consider only frequencies $\omega$ much smaller than the self-resonance frequency $\omega_r \sim c / W$ of the coil, allowing us to neglect the capacitance term. 
With these simplifying assumptions, the current is uniform along the wire, and the $Q$-factor \added{is defined as the ratio of the stored magnetic energy to the magnetic losses. For the purpose of this paper, an equivalent and more convenient definition of $Q$ is} the ratio of the imaginary and real parts of the complex impedance $Z = R + \text{i} \omega L$:
\begin{equation}
Q(\omega) = \frac{\im Z}{\re Z} = \frac{\omega L(\omega)}{R(\omega)}.
\label{eqn:Q_def}
\end{equation}
Because of induced eddy currents, $R(\omega)$ is coil-shape dependent,
so that the competition between the inductance and the losses poses a nontrivial optimization problem for $Q(\omega)$.

\begin{figure}[t]
\includegraphics[width = 2.50in]{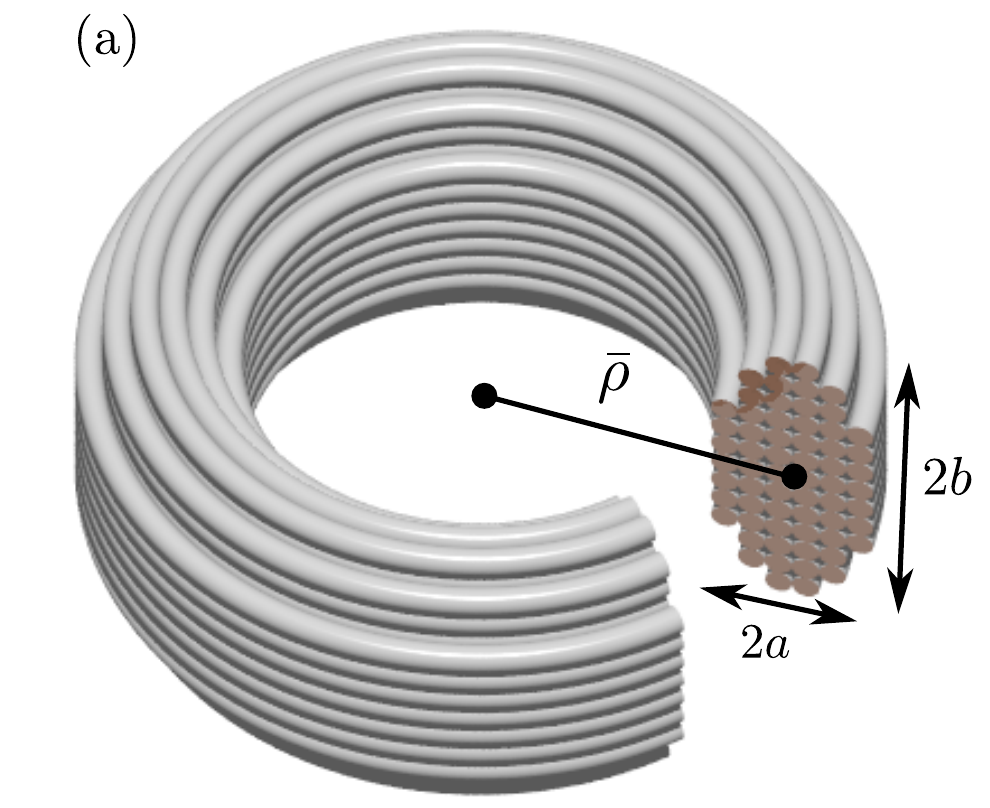}
\includegraphics[width =2.50in]{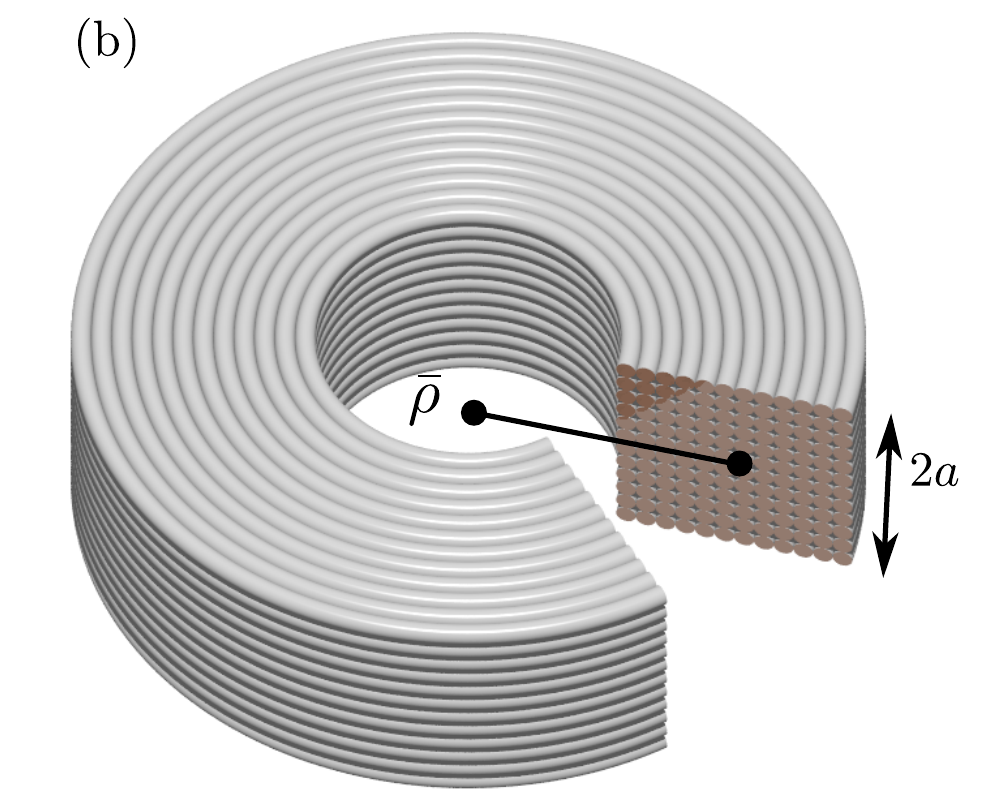}
\caption{Schematics of multi-layer coils with (a) elliptic and (b) square cross-sections. 
}
\label{fig:ind_shapes}
\end{figure}

Our electrodynamic problem has roots in a magnetostatic problem first studied by Gauss.~\cite{Gauss1867}
Specifically, in the limit $\omega \to 0$, the
effective resistance $R$ approaches the dc resistance $R(0) = 4 W / (\pi \sigma d_i^2)$,
where $\sigma$ is the core conductivity,
so that maximizing $Q$ is equivalent to maximizing $L$.
Gauss assumed that the coil of the highest $L$ under the aforesaid constraints is a toroidally wound solenoid with a nearly circular cross-section, Fig.~\ref{fig:ind_shapes}(a).
Later, Maxwell~\cite{Maxwell1892} revisited the problem and treated a more practical case of a square cross-section, Fig.~\ref{fig:ind_shapes}(b).
Maxwell's analysis was improved by Rosa and Grover.~\cite{Rosa1912}
Building on their work, Brooks proposed that
the mean radius of the optimal coil is approximately $3 / 2$ of the side of the square.~\cite{Brooks1931}
The inductance of this coil is $0.656 L_c$,
where
\begin{equation}
L_c = \frac{\mu_0}{4\pi} \frac{W^{5/3}}{d^{\,2/3}}.
\label{eqn:L_c}
\end{equation}
Optimization of inductors with nonmagnetic cores became topical again
in the 1970's when toroidal coils (wound in the poloidal direction) were
brought in a wider use in plasma physics and energy storage research.
The case of a single-layer toroid was solved by Shafranov.~\cite{Shafranov1973, Gralnick1976}
Multilayer coils were studied by Murgatroyd~\cite{Murgatroyd1989, Murgatroyd2000} who found that the inductance of the optimal toroid is $0.29 L_c$.
The reduction compared to the Brooks coil is presumably because the toroid generates no stray magnetic field.
Murgatroyd reviewed the $5 / 3$ power-law of \deleted{Eq.}~\eqref{eqn:L_c} and other 
properties of optimal inductors in his excellent summary.~\cite{Murgatroyd1989}
For example, the characteristic size of such inductors is set by
\begin{equation}
\rho_c = \frac12 ({W d^2})^{1 / 3}.
\label{eqn:barrho}
\end{equation}
\added{These scaling laws apply assuming the wire bundle forming the cross-section of the coil can be approximated by a continuum current distribution, which is legitimate if the number of turns $N$ is large enough. For example, the relative error in the following inductance calculation due to this approximation scales as $N^{-1/2}$, as shown by Maxwell~\cite{Maxwell1892}. Adopting this continuum approach,} below we derive scaling laws for finite-$\omega$ optimal inductors
in terms of two additional characteristic scales:
\begin{equation}
\omega_c \equiv \frac{8\pi}{Q_c} \frac{d^2}{\mu_0 \sigma d_i^4},
\qquad
Q_c \equiv \frac{2 \rho_c}{d_i}.
\label{eqn:omega_c}
\end{equation}
The former is the frequency at which the eddy-current losses become comparable with the dc Ohmic ones,
the latter is the order of magnitude of the $Q$-factor at $\omega_c$.

\section*{Low frequencies}
\begin{figure}[bt]
\begin{center}
\includegraphics[width = 5.0in]{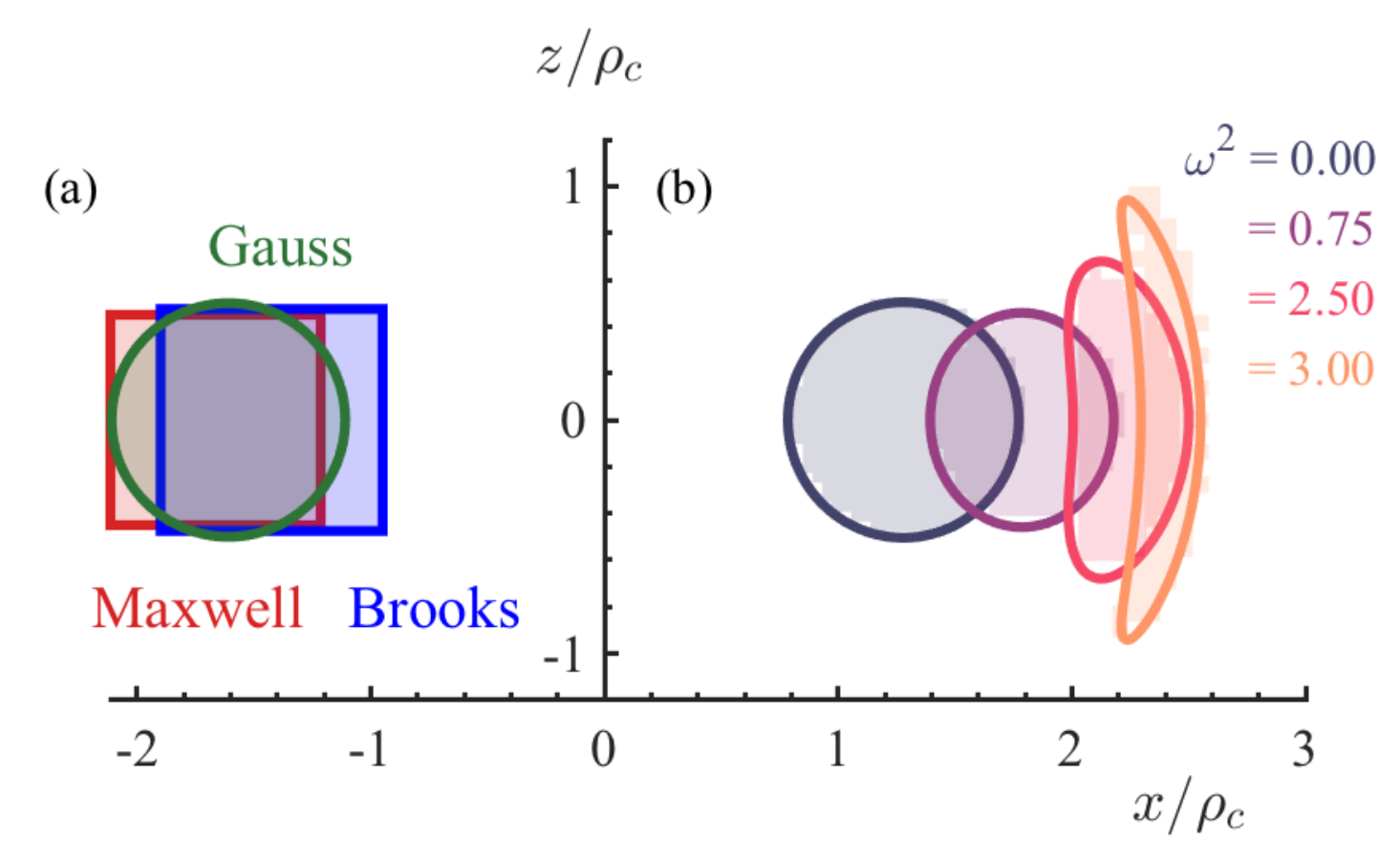}
\end{center}
\caption{Cross-sections of the optimal coils.
(a) Designs proposed by Gauss,~\cite{Gauss1867} Maxwell,~\cite{Maxwell1892} and Brooks.~\cite{Brooks1931}
(b) Results obtained in this work. 
The cross-section evolves from near-circular to elliptic to sickle-shaped
as $\omega$ increases.
The shading represent the local wire density $n(\mathbf{r})$ computed on a $30\times 30$ grid.
The curves serve as guides to the eye.
The wire density is seen to switch from $0$ to $n_2$
with few or no intermediate values.
The numbers on the axes are $x$ and $z$ coordinates in units of $\rho_c$.
The legend indicates the magnitudes of $(\omega / \omega_c)^2$.
}
\label{fig:evolution}
\end{figure}

We begin with answering Gauss' question about the dc inductance.
It was posed by him 150 years ago but apparently has not been settled yet.
Gauss' calculation can be summarized as follows.
An estimate of $L$ is provided by the approximate formula~\cite{Maxwell1892}
\begin{equation}
L = \mu_0 N^2 \bar\rho
\left[\ln\left(\frac{8 \bar\rho}{{\mathrm{\scriptstyle GMD}}}\right) - 2\right],
\label{eqn:L_M}
\end{equation}
where $N$ is the total number of turns in the coil and $\bar \rho$ is their mean radius.
Note that $2\pi \bar\rho N = W$.
Parameter ${\mathrm{\scriptstyle GMD}}$ is the geometric mean distance.
In the continuum limit, appropriate for large $N$, it is defined via
\begin{equation}
\ln ({\mathrm{\scriptstyle GMD}}) = \frac{1}{A^2}
\iint\, \ln |\mathbf{r} - \mathbf{r}'| {d^2 r d^2 r'},
\label{eqn:GMD}
\end{equation}
where positions $\mathbf{r} = (\rho, z)$, $\mathbf{r}' = (\rho', z')$ vary over the cross-section of the coil,
of the area $A = N / n_2$.
According to \deleted{Eq.}~\eqref{eqn:L_M},
to maximize $L$ for a given $N$ (or $\bar\rho$) we need to minimize ${\mathrm{\scriptstyle GMD}}$ at fixed $A$.
It can be proven~\cite{Polya1951} that the solution is a circle of radius $a = \sqrt{A / \pi}$
whose ${\mathrm{\scriptstyle GMD}}$ is~\cite{Rosa1912} $\text{e}^{-1 / 4} a$.
Minimizing $L$ with respect to $\bar\rho / a$,
Gauss obtained $\bar\rho / a = \text{e}^{13 / 4} / 8 = 3.22$.
Such a mean-radius to half-height ratio is noticeably different from either $3.7$ or $3$ advocated by, respectively, Maxwell and Brooks, see Fig.~\ref{fig:evolution}(a),
suggesting that this method is too crude to reveal the true optimal coil geometry.

%

To glean a more accurate answer, we tackled the problem numerically.
We expressed the inductance and the wire-length constraint in the form of integrals,
\begin{align}
L &= \iint \,n(\mathbf{r}) M(\mathbf{r}, \mathbf{r}') n(\mathbf{r}')\, d^2r\, d^2r',
\label{eqn:L_num_model}\\
W &= \int \,n(\mathbf{r})\, 2\pi\! \rho\,  d^2r,
\label{eqn:W_num_model}
\end{align}
where $0 \leq n(\mathbf{r}) \leq n_2$ is the number of turns per unit area at position $\mathbf{r}$. Function $M(\mathbf{r}, \mathbf{r}')$, given by
\begin{equation}
\begin{split}
M(\mathbf{r}, \mathbf{r}') &= \mu_0 \sqrt{\frac{\rho \rho'}{m}}
\left[(2 - m) K(m) - 2 E(m)\right],
\\
m &= \frac{1}{1 + k^2},\quad k = \frac{|\mathbf{r} - \mathbf{r}'|}{\sqrt{4 \rho \rho'\vphantom{|}}}
\end{split}
\label{eqn:M}
\end{equation}
is the mutual inductance of co-axial \replaced{line currents}
{ring currents}~\cite{Rosa1912} piercing the cross-section at $\mathbf{r}$ and $\mathbf{r}'$; $K(m)$ and $E(m)$ are the complete elliptic integrals.
We approximated the integrals in \deleted{Eqs.}~\eqref{eqn:L_num_model}, \eqref{eqn:W_num_model} by sums over a finite two-dimensional grid
and performed the constrained maximization of $L$ numerically. 
The mean radius of the optimal coil is $\bar\rho = 1.28 \rho_c$.
The cross-section of the coil is not a circle;
it is better approximated by an ellipse of dimensions
\begin{equation}
\xi_1 \equiv \frac{\bar{\rho}}{a} = 2.54,
\qquad
\xi_2 \equiv \frac{\bar{\rho}}{b} = 2.61,
\label{eqn:a_and_b}
\end{equation}
represented by the curve labeled $\omega^2 = 0$ in Fig.~\ref{fig:evolution}(b).
The cross-section is fully packed, so that
\begin{equation}
n(\mathbf{r}) = n_2\,
\Theta\left(1 - \frac{(\rho - \bar{\rho})^2}{a^2} - \frac{z^2}{b^2}\right),
\label{eqn:ellipse}
\end{equation}
where $\Theta(x)$ is the unit step-function.
Finally, the coil inductance is
\begin{equation}
L = 0.663 L_c,
\label{eqn:L}
\end{equation}
which is $1\%$ larger than that of the Brooks coil.

Encouraged by the simplicity of these results,
we rederived them as follows.
We started with the expansion~\cite{Rosa1912}
\begin{equation}
M(\mathbf{r}, \mathbf{r}') \simeq \mu_0\sqrt{\rho \rho'}
\left[
\left(1 + \frac{3 k^2}{4} \right)
\ln \frac{4}{k}
-2 - \frac{3 k^2}{4}
\right],
\label{eqn:M_expansion}
\end{equation}
valid for $k \ll 1$ [Eq.~\eqref{eqn:M}],
and evaluated the integral in \deleted{Eq.}~\eqref{eqn:L_num_model} analytically
for the elliptic cross-section defined by \deleted{Eq.}~\eqref{eqn:ellipse}.
The result can be written as
\begin{align}
L &= \mu_0 N^2 \bar{\rho}\, \Lambda,
\label{eqn:L_elliptic}
\\
\Lambda &=
\left(1 + \frac{1}{32} \frac{\xi_2^{2} + 3 \xi_1^{2}}{\xi_1^{2} \xi_2^{2}}\right)
\ln \left(\frac{16\, \xi_1 \xi_2}{\xi_1 + \xi_2}\right) - \frac{7}{4} 
\notag\\
&+ \frac{7}{96} \frac{1}{\xi_1^2}
+ \frac{1}{32} \frac{\xi_2^{2} - 3 \xi_1^{2}}{\xi_1^{2} \xi_2^{2}} \frac{\xi_1}{\xi_1 + \xi_2},
\label{eqn:Lambda_elliptic}
\end{align}
which is a generalization of Rayleigh's formula~\cite{Rayleigh1912} for the $b = a$ case
and a key improvement over \deleted{Eq.}~\eqref{eqn:L_M}.
Using this formula for $L$ and another one, $W = \pi a b \bar{\rho} n_2$,
for the length constraint,
we were able to easily solve for the optimal $\xi_1$, $ \xi_2$ numerically, reproducing \deleted{Eq.}~\eqref{eqn:a_and_b}.

\begin{figure}[th]
	\begin{center}
		\includegraphics[width=4.0in]{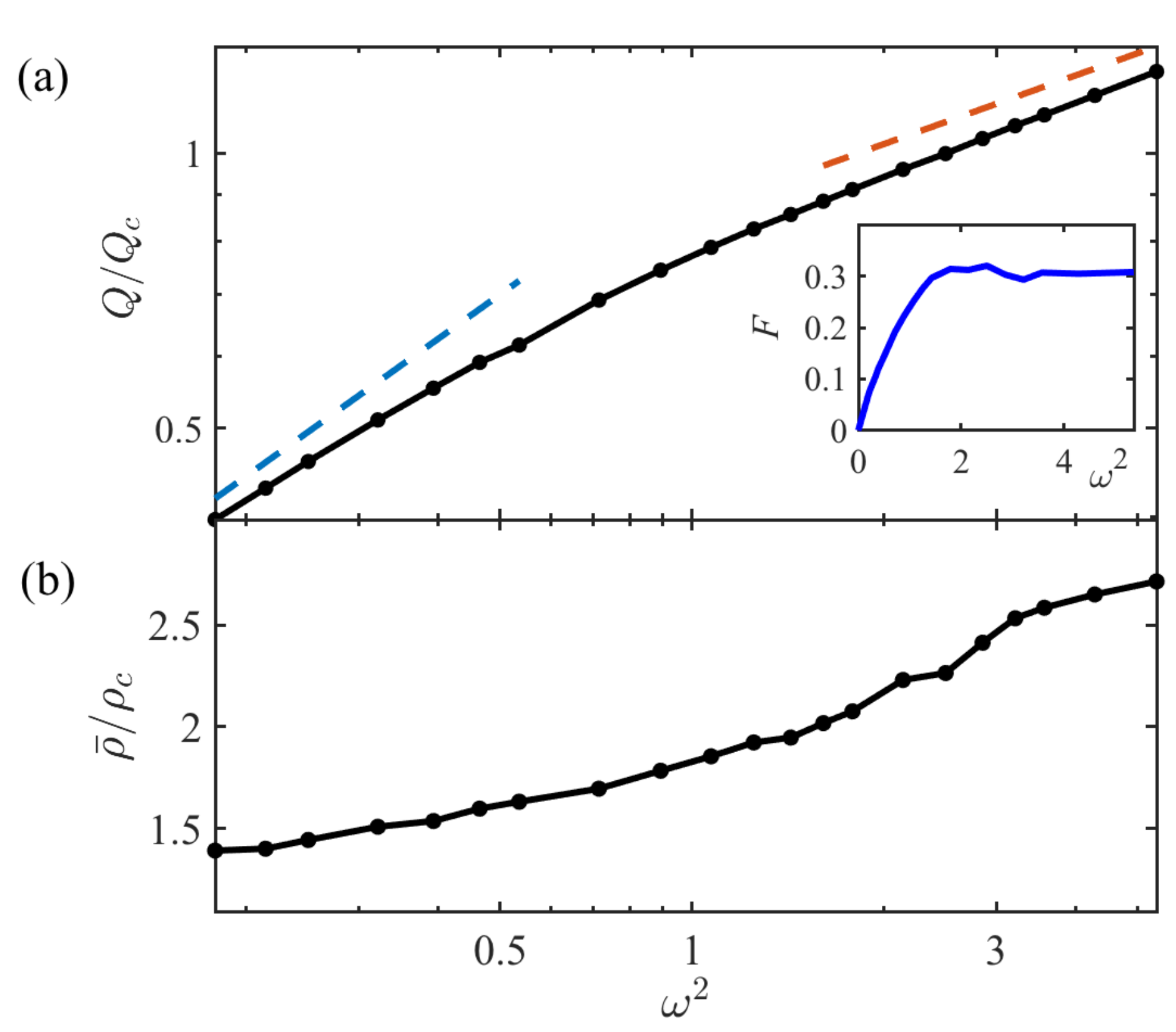}
	\end{center}
	\caption{(a) $Q$-factor of the optimal coil as a function of $\omega^2 / \omega_c^2$.
		The connected dots are our numerical results.
		The two dashed lines indicate the expected low- and intermediate-frequency scaling.
		Inset: loss factor $F$ \textit{vs}. $\omega^2 / \omega_c^2$.
		(b) Mean radius $\bar{\rho}$ of the coil in units of $\rho_c$
		as a function of $\omega^2 / \omega_c^2$.
	}
	\label{fig:Q_phase}
\end{figure}

Returning to the $Q$-factor,
we rewrite \deleted{Eq.}~\eqref{eqn:Q_def} in terms of our characteristic scales $L_c$, $Q_c$, $\omega_c$:
\begin{equation}
Q = \frac{\pi}{2} \frac{\omega}{\omega_c} \frac{L / L_c}{1 + F(\omega)} Q_c,
\label{eqn:Q_def2}
\end{equation}
where we introduced the loss enhancement factor
\begin{equation}
F(\omega) \equiv \frac{R(\omega)}{R(0)}  - 1.
\label{eqn:R}
\end{equation}
Below we show that at low frequencies $\omega \ll \omega_c$, the loss factor behaves as
\begin{equation}
F(\omega) = 0.305\, \frac{\omega^2}{\omega_c^2}\,.
\label{eqn:F_lowfreq}
\end{equation}
At such frequencies, $F \ll 1$ is negligible, $L$ is virtually unchanged from the dc value, and so the $Q$-factor is linear in $\omega$:
\begin{equation}
\frac{Q}{Q_c} = 1.04\, \frac{\omega}{\omega_c},
\quad \omega \ll \omega_c,
\label{eqn:Q_lowfreq_W}
\end{equation}
see Fig.~\ref{fig:Q_phase}.

\section*{Proximity effect losses}
The finite-frequency losses in coils are traditionally attributed to the combination of the skin and proximity effects.~\cite{Butterworth1925}
The latter, due to the collective field $\mathbf{H}(\mathbf{r})$ of all the turns of the wire,
dominates in multi-layer coils of interest to us if $\omega$ is not too high, such that $\delta \gg d_i$,
where
\begin{equation}
\delta(\omega) = \sqrt{\frac{\vphantom{|}2}{\mu_0 \omega \sigma}}
\label{eqn:delta}
\end{equation}
is the skin depth.
Under the stated condition of weak skin effect, the loss factor takes the form
\begin{equation}
F(\omega) = \frac{\pi^2}{64 W}
\frac{d_i^6}{\delta^4}
\int \,\mathbf{H}^2(\mathbf{r}) n(\mathbf{r})\, 2\pi\! \rho\,  d^2r
\label{eqn:F_num_model}
\end{equation}
where $\mathbf{H}$, equal to the curl of a vector potential, is
\begin{equation}
\mathbf{H}(\mathbf{r}) = \frac{1}{2\pi\mu_0\rho} (\hat{\mathbf{z}} \partial_\rho - \hat{\bm{\rho}} \partial_z)
\int \,M(\mathbf{r}, \mathbf{r}') n(\mathbf{r}') \text{d}^2r'.
\label{eqn:H_num_model}
\end{equation}
In general, these expressions have to be evaluated numerically.
However, we can estimate $F$ analytically for a coil with the elliptic cross-section,
\deleted{Eq.}~\eqref{eqn:ellipse}.
Retaining only the leading-order terms in $k \sim \max(a, b) / \bar{\rho} \ll 1$
in \deleted{Eq.}~\eqref{eqn:M_expansion},
we find
\begin{equation}
\mathbf{H}(\mathbf{r}) =
\frac{n_2}{a + b} \left[
a z\, \hat{\bm{\rho}} - (\rho - \bar{\rho}) b\, \hat{\mathbf{z}}
\right].
\label{eqn:H_ellipse}
\end{equation}
Substituting this into \deleted{Eq.}~\eqref{eqn:F_num_model}, we get
\begin{equation}
F = \frac{1}{8} \left(\frac{d_i^3}{\delta^2 d^2}
\frac{a b}{a + b}\right)^2
 = \frac{\pi^2}{2}\, \frac{\omega^2}{\omega_c^2}
 \left(\frac{1}{\rho_c}\, \frac{a b}{a + b}\right)^2,
\label{eqn:F_ellipse}
\end{equation}
which is a generalization of Howe's formula for a multi-stranded round wire.~\cite{Howe1917}
Finally, using \deleted{Eqs.}~\eqref{eqn:barrho} and \eqref{eqn:a_and_b}, we arrive at \deleted{Eq.}~\eqref{eqn:F_lowfreq}.
At the border of its validity, $\omega \approx \omega_c$,
that equation predicts $F \approx 0.3$
assuming the wire is long enough so that $\delta / d_i \approx 0.2\, (W/d)^{1/6} \gg 1$.
%

\section*{Intermediate frequencies}
At $\omega \gg \omega_c$ the competition between inductance and proximity losses
is expected to cause flattening of the cross-section of the optimal coil.
We confirmed this hypothesis by numerical simulations
based on \deleted{Eqs.}~\eqref{eqn:L_num_model}, \eqref{eqn:W_num_model},
\eqref{eqn:F_num_model}, and \eqref{eqn:H_num_model}.
Our results for a few representative $\omega$ are shown in Fig.~\ref{fig:evolution}(b).
As frequency increases, the cross-section first becomes oval and then sickle-shaped.
Figure~\ref{fig:Q_phase} presents the $Q$-factor and the mean radius $\bar{\rho}$ obtained from these simulations.
The plot in the main panel of Fig.~\ref{fig:Q_phase}(a) suggests that the linear scaling of $Q(\omega)$ changes
to a square-root law above the frequency $\omega_c$ as the cross-section begins to flatten and bend. 
The inset of Fig.~\ref{fig:Q_phase}(a) illustrates that
the loss factor grows as predicted by \deleted{Eq.}~\eqref{eqn:F_lowfreq} at $\omega / \omega_c < 1$
but reaches a constant $F \approx 0.3$ at $\omega / \omega_c > 1$.

We can shed light on the observed $\omega / \omega_c > 1$ behaviors using our elliptical cross-section model.
Assuming $a\ll b$, we derive the following analytical expressions for $a$ and $b$ in terms
of dimensionless parameters $\xi_2 = \bar{\rho} / b$ and $F$:
\begin{equation}
\frac{a}{\rho_c} = \sqrt{\frac{F}{2\pi^2}}\, \frac{\omega_c}{\omega}\,,
\qquad
\frac{b}{\rho_c} = \sqrt{\frac{1}{\pi \xi_2} \frac{\rho_c}{a}}\,.
\label{eqn:a_b_intfreq}
\end{equation}
They entail that $Q$ at a given $\omega$ has the scaling form
\begin{equation}
Q(\xi_2, F) = \frac{F^{1 / 4}}{1 + F}\, q\left(\xi_2\right).
\label{eqn:Q_intfreq1}
\end{equation}
Hence, $Q$ at fixed $\xi_2$ reaches its maximum at $F = 1/3$,
which is close to our numerical result. 
Freezing $F$ at $1 / 3$ and
maximizing $Q$ with respect to $\xi_2$, we arrived at
\begin{align}
\frac{Q}{Q_c} &= 0.85\, \sqrt{\frac{\omega}{\omega_c}}\,,
& \frac{\bar{\rho}}{\rho_c} &= 1.6\, \sqrt{\frac{\omega}{\omega_c}}\,,
\label{eqn:Q_and_rho_intfreq}\\
\frac{a}{\rho_c} &= 0.26\, \frac{\omega_c}{\omega},
& \xi_2 &= 2.13.
\label{eqn:a_and_xi_2}
\end{align}
The first equation in \deleted{Eq.}~\eqref{eqn:Q_and_rho_intfreq},
represented by the upper dashed line in Fig.~\ref{fig:Q_phase}(a),
is within $10\%$ from the simulation results.
The second equation in \deleted{Eq.}~\eqref{eqn:Q_and_rho_intfreq},
has a similar level of agreement with the data in Fig.~\ref{fig:Q_phase}(b).
This is satisfactory considering that $\omega / \omega_c$ is not truly large and that
our analytical model is oversimplified.

\section*{High frequencies}
From now on we focus on the practical case of densely packed,
thinly insulated wires, $d_i \approx d$.
Per \deleted{Eqs.}~\eqref{eqn:delta} and \eqref{eqn:a_b_intfreq},
at frequency $\omega_s = \omega_c Q_c / (2\pi) \gg \omega_c$ both the width $2a$ of the thickest part of the winding and the skin depth $\delta$ become of the order of $d$.
This implies that at $\omega \gg \omega_s$ the optimal coil is (i) single-layered
and (ii) strongly affected by the skin effect.
In view of the former,
we can fully specify the cross-sectional shape of the coil by a function $\rho(z)$
and replace \deleted{Eqs.}~\eqref{eqn:L_num_model} and \eqref{eqn:W_num_model} by
\begin{align}
L &= n_1^2 \iint\! \, M(\mathbf{r}, \mathbf{r}')
ds\, ds',
\label{eqn:L_num_model2}\\
W &= n_1 \int\! \,2\pi \rho(z) ds,
\quad
\text{d} s = \sqrt{1 + \rho^{\prime 2}(z)}\, dz,
\label{eqn:W_num_model2}
\end{align}
with $n_1 \sim 1 / d$ being the number of turns per unit arc length $s$ of the cross-section.
Equation~\eqref{eqn:F_num_model} gets modified as well.
As first shown by Rayleigh,~\cite{Rayleigh1886} a single straight round wire is characterized by the loss factor $F_s = {d_i} / ({4\delta}) \gg 1$,
due to confinement of the current to a $\delta$-thick skin layer at the conductor surface.
In a coil or in a bunch of parallel wires,
inter-wire interactions cause further nonuniformity of the current in the skin layer.
As a result, the loss factor increases beyond Rayleigh's $F_s$:
\begin{equation}
\frac{F}{F_s} = \lambda
+ \frac{d_i^2 n_1}{8 W} \int\, \left[f {H}^2_{\parallel}(z) + g {H}^2_{\perp}(z)\right]
 2\pi\! \rho\, ds,
\label{eqn:F_num_model2}
\end{equation}
where ${H}_{\parallel}(z)$ and ${H}_{\perp}(z)$ are the components of $\mathbf{H}(\mathbf{r})$
parallel and perpendicular to the layer,

%
\begin{figure}[t]
	\begin{center}
		\includegraphics[width=4.0in]{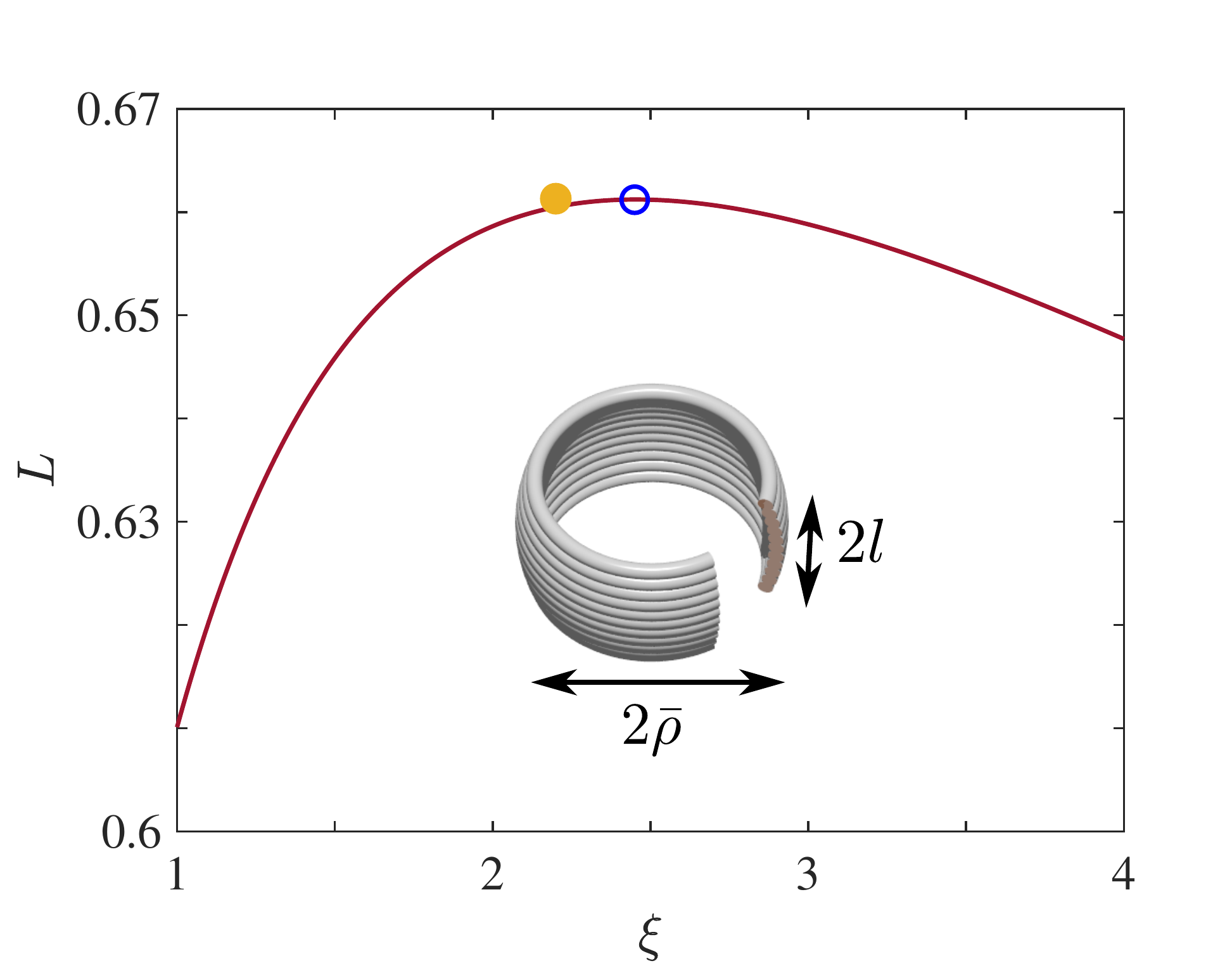}
	\end{center}
	\caption{Inductance $L$ of a constant-radius single-layer coil as a function of $\xi = \bar{\rho} / l$.
		The open dot labels the maximum on the curve.
		The filled dot shows the true optimum.
		$L$ is in units of $\mu_0 W^{3 / 2} / (2\pi \sqrt{d})$.
		Inset: definitions of $\bar{\rho}$ and $l$.
	}
	\label{fig:L_solenoid}
\end{figure}

\begin{equation}
{H}_{\parallel}(z) = \frac{H_\rho \rho' + H_z}{\sqrt{1 + \rho^{\prime 2}}},
\qquad
{H}_{\perp}(z) = \frac{H_\rho - H_z \rho'}{\sqrt{1 + \rho^{\prime 2}}}.
\label{eqn:H_parallel_perp}
\end{equation}
The dimensionless coefficients $\lambda$, $f$, and $g$
introduced by Butterworth~\cite{Butterworth1925} depend on the wire packing density $n_1 d_i$
and have to be calculated numerically.~\cite{Smith1972}
The optimization of $Q$ using the entire set of these complicated equations appears to be challenging, so we have not attempted it.
On the other hand, the solution for $\rho(z)$ we present below is a
nearly constant function.
For such functions the loss factor $F$ should be weakly shape dependent, in which case to maximize $Q$ it is sufficient to maximize $L$ alone.
We accomplished the latter numerically
using \deleted{Eqs.}~\eqref{eqn:L_num_model2} and \eqref{eqn:W_num_model2},
in which we additionally dropped the $\sqrt{1 + \rho^{\prime 2}}$ factors.
The optimal solenoid shape we found is slightly convex,
as depicted schematically in Fig.~\ref{fig:L_solenoid}, with the aspect ratio $\xi = \bar{\rho} / l = 2.20$ and curvature $0.0024 / l$.
Note that $\xi$ is numerically close to $\xi_2$ in the intermediate frequency regime,
\deleted{Eq.}~\eqref{eqn:a_and_xi_2}.
Substituting the obtained $L$ into \deleted{Eq.}~\eqref{eqn:Q_def2}, we got
\begin{equation}
\frac{Q}{Q_c} = \frac{2.34}{F / F_s}\, \sqrt{\frac{\omega}{\omega_c}}\,,
\quad
\omega \gg \omega_s,
\label{eqn:Q_solenoid}
\end{equation}
which is similar to \deleted{Eq.}~\eqref{eqn:Q_and_rho_intfreq} but has a different coefficient.
This type of high-frequency behavior is actually well-known in radio engineering.~\cite{Medhurst1947a, Medhurst1947b}

In an effort to rederive these results more simply,
we considered a family of constant-radius solenoids
whose inductance is given by Lorenz's formula~\cite{Rosa1912}
\begin{align}
L = \frac{8}{3} \mu_0 n_1^2 \rho^3 \left[\frac{2m - 1}{m \sqrt{m}} E(m) + \frac{1 - m}{m \sqrt{m}} K(m) - 1\right],
\label{eqn:l_solenoid}
\end{align}
where $m = {\rho^2} / ({\rho^2 + l^2})$.
As seen in Fig.~\ref{fig:L_solenoid},
the maximization of this $L$
(under the constraint $4\pi \rho l n_1 = W$)
gives
$L = 0.661 \mu_0 W^{3 / 2} / (2\pi \sqrt{d}\,)$,
in agreement with Murgatroyd.~\cite{Murgatroyd1989}
This value of the inductance is only $\sim 1\%$ lower than the true optimum, \added{which corresponds to the slightly convex shape we found here}.
Yet the best aspect ratio for the constant-radius solenoid proves to be $2.46$,
a $13\%$ larger than for our optimal coil. 

\section*{Discussion}
In this work we studied theoretically the highest possible $Q$-factor of an inductor
wound from a given piece of wire.
Real inductors used in various practical applications~\cite{Burghartz2003, Tumanski2007, Karalis2008, Bahl2001} are made under numerous additional constraints, such as minimal cost, \deleted{or} ease of manufacturing, \added{or current handling capacity.}
Depending on the application, a multitude of related optimization problems may arise. 
Our calculation provides a fundamental upper bound on $Q$ and
its scaling with wire length, diameter, and frequency.
At the highest frequencies we considered, $Q(\omega)$ grows according to the square-root law.
We expect this law to persist until either capacitance effects or radiative losses or frequency dispersion of $\sigma$ neglected in our theory become important.
For example, the capacitance effects restrict the validity of \deleted{Eq}.~\eqref{eqn:Q_solenoid} to frequencies below the self-resonance frequency
$\omega_r \sim c / W$.
Hence, this equation may apply only if $\omega_r / \omega_s \sim Z_0 / R(0) \gg 1$,
i.e., if the dc resistance of the wire is much smaller than
the impedance of free space $Z_0 = c \mu_0 = 377\,\Omega$.
However, if $R(0)$ is too low, then \added{dipole radiation}\cite{Smith, Landau1975} losses, growing as $\omega^4$,
could surpass the Ohmic ones.
Consideration of these additional physical effects
is relevant for optimizing inductors used in resonators, antennas, and metamaterials,
and so it could be an interesting topic for future research. 

\section*{Methods} \added{To optimize inductance, the equations \eqref{eqn:L_num_model} and \eqref{eqn:W_num_model} were replaced by sums:}
	\begin{align}
	&L[n] = \sum_i \sum_jn_i M_{ij} n_j\\
	& W[n] = \sum_i2\pi\rho_in_i
	\label{eqn:method_L}
	\end{align}
\added{over finite two-dimensional grid of points $\mathbf{r}_i$ in the cross-section of the coil . The off-diagonal elements $M_{ij} = M(\mathbf{r}_i, \mathbf{r}_j)$ were found from \eqref{eqn:M}. 
To find the diagonal elements of the matrix, the mutual inductance of two rings of radius $\rho_i$ offset by a vertical distance, $M(\mathbf{r_i}, \mathbf{r_i} + a_\mathrm{GMD}\,\mathbf{\hat{z})}  )$, was computed, which is a good approximation to the self-inductance of a thin wire~\cite{Grover1946}. 
The matrix $M$ is positive definite, and so the maximization of $L$ is a convex constrained optimization problem, which we solved using MATLAB's\cite{MATLAB2018} built-in \texttt{quadprog} function, yielding the optimal distribution of currents $n_i$. 

The full problem including the proximity and skin effects is more complicated; to begin with, it is no longer obviously convex due to the additional factor in~\eqref{eqn:F_num_model}. The magnetic field entering this equation can in principle be found from\eqref{eqn:H_num_model}. We used an equivalent method, as follows. For each point on a two-dimensional grid, the magnetic field vector at coordinate $i$ due to the current at coordinate $j$ was calculated using the well-known formula\cite{Simpsons2001} for the magnetic field of a ring current, yielding the matrix $H_{ij}$. The loss factor was then calculated using the discretized version of \eqref{eqn:F_num_model},}

\begin{equation}
F[n] = \frac{\pi^3}{32 W}
\frac{d_i^6}{\delta^4}
 \sum_j\left(\sum_iH_{ij} n_i\right)^2\rho_j.
 \label{eqn:method_F}
\end{equation}
\added{The optimal distribution of current $n_i$ was then obtained by maximizing}

\begin{equation}
Q[n] = \frac{L[n]}{1 + F[n]},
\label{eqn_method_Q}
\end{equation} 
\added{numerically, using the built-in \texttt{fmincon} function in MATLAB. 
This function requires an initial guess, which we chose to be random. We verified that the results of the optimization were independent of the starting values of $n_i$ and satisfied the constraint \eqref{eqn:method_L} up to the specified tolerance of $10^{-6}$. For guiding the eye, the values $n_i$ depicted in Figure~\ref{fig:evolution} were supplemented with smooth envelope curves. For the two lower frequencies, ellipses were used, and for the higher two,  sickle-shaped curves satisfying $c_0 = (x^2 + y^2 - c_1)^2 + c_2 y^2$ were used with suitable choices of $c_0, c_1, c_2$. }

\section*{Acknowledgments}
This work was supported by The Office of Naval Research under grant N00014-18-1-2722, by the US Department of Energy under grant DE-SC0018218 (apropos energy storage), and by General ElectroDynamics International, Inc.
We thank Yu. A. Dreizin for discussions that inspired this study and 
also B. I. Shklovskii and E. Yablonovitch for comments on the manuscript.
 
 \section*{Author contributions}
\replaced{Both authors contributed to the calculations presented in this work and writing the manuscript. }{Both authors contributed to the calculations presented in this work. A. R. performed the numerical optimizations. M. M. F. proposed and supervised the study. Both authors wrote and reviewed the manuscript. }
 
 \section*{Competing interests}
 The authors declare no competing interests. 
 
\def\bibsection{\section*{\refname}} 

\bibliography{Optimal_Q}

\begin{thebibliography}{10}
\urlstyle{rm}
\expandafter\ifx\csname url\endcsname\relax
  \def\url#1{\texttt{#1}}\fi
\expandafter\ifx\csname urlprefix\endcsname\relax\def\urlprefix{URL }\fi
\expandafter\ifx\csname doiprefix\endcsname\relax\def\doiprefix{DOI: }\fi
\providecommand{\bibinfo}[2]{#2}
\providecommand{\eprint}[2][]{\url{#2}}

\bibitem{Medhurst1947a}
\bibinfo{author}{Medhurst, R.~G.}
\newblock \bibinfo{journal}{\bibinfo{title}{{H}. {F}. {R}esistance and
  self-capacitance of single-layer solenoids}}.
\newblock {\emph{\JournalTitle{{W}ireless {E}ng.}}}
  \textbf{\bibinfo{volume}{24}}, \bibinfo{pages}{35--43},
  \doiprefix\url{https://worldradiohistory.com/UK/Experimental-Wireless/40s/Wireless-Engineer-1947-03.pdf}
  (\bibinfo{year}{1947}).

\bibitem{Medhurst1947b}
\bibinfo{author}{Medhurst, R.~G.}
\newblock \bibinfo{journal}{\bibinfo{title}{{H}. {F}. {R}esistance and
  self-capacitance of single-layer solenoids}}.
\newblock {\emph{\JournalTitle{{W}ireless {E}ng.}}}
  \textbf{\bibinfo{volume}{24}}, \bibinfo{pages}{82--90},
  \doiprefix\url{https://worldradiohistory.com/UK/Experimental-Wireless/40s/Wireless-Engineer-1947-03.pdf}
  (\bibinfo{year}{1947}).

\bibitem{Gauss1867}
\bibinfo{author}{Gauss, C.~F.}
\newblock \emph{\bibinfo{title}{Werke}}, vol.~\bibinfo{volume}{5}
  (\bibinfo{publisher}{Cambridge University Press},
  \bibinfo{address}{Cambridge}, \bibinfo{year}{2011}).
\newblock \bibinfo{note}{(first published in 1867)}.

\bibitem{Maxwell1892}
\bibinfo{author}{Maxwell, J.~C.}
\newblock \emph{\bibinfo{title}{Treatise on {E}lectricity and {M}agnetism}},
  vol.~\bibinfo{volume}{2} (\bibinfo{publisher}{Dover}, \bibinfo{address}{New
  York}, \bibinfo{year}{1954}), \bibinfo{edition}{third} edn.
\newblock \bibinfo{note}{, article 706, article 693}.

\bibitem{Rosa1912}
\bibinfo{author}{Rosa, E.~B.} \& \bibinfo{author}{Grover, F.~W.}
\newblock \bibinfo{journal}{\bibinfo{title}{Formulas and tables for the
  calculation of mutual and self inductance. ({R}evised.)}}.
\newblock {\emph{\JournalTitle{Bulletin of the Bureau of Standards}}}
  \textbf{\bibinfo{volume}{8}}, \bibinfo{pages}{1--237},
  \doiprefix\url{http://doi.org/10.6028/bulletin.185} (\bibinfo{year}{1912}).

\bibitem{Brooks1931}
\bibinfo{author}{Brooks, H.~B.}
\newblock \bibinfo{journal}{\bibinfo{title}{Design of standards of inductance,
  and the proposed use of model reactors in the design of air-core and
  iron-core reactors}}.
\newblock {\emph{\JournalTitle{Bureau of Standards Journal of Research}}}
  \textbf{\bibinfo{volume}{7}}, \bibinfo{pages}{298--328},
  \doiprefix\url{http://dx.doi.org/10.6028/jres.007.016}
  (\bibinfo{year}{1931}).

\bibitem{Shafranov1973}
\bibinfo{author}{Shafranov, V.~D.}
\newblock \bibinfo{journal}{\bibinfo{title}{Optimal shape of a toroidal
  solenoid}}.
\newblock {\emph{\JournalTitle{Sov. Phys. - Tech. Phys.}}}
  \textbf{\bibinfo{volume}{17}}, \bibinfo{pages}{1433--1437}
  (\bibinfo{year}{1973}).

\bibitem{Gralnick1976}
\bibinfo{author}{Gralnick, S.~L.} \& \bibinfo{author}{Tenney, F.~H.}
\newblock \bibinfo{journal}{\bibinfo{title}{Analytic solutions for
  constant‐tension coil shapes}}.
\newblock {\emph{\JournalTitle{{J}. {A}ppl. {P}hys.}}}
  \textbf{\bibinfo{volume}{47}}, \bibinfo{pages}{2710--2715},
  \doiprefix\url{http://doi.org/10.1063/1.322993} (\bibinfo{year}{1976}).

\bibitem{Murgatroyd1989}
\bibinfo{author}{Murgatroyd, P.~N.}
\newblock \bibinfo{journal}{\bibinfo{title}{The optimal form for coreless
  inductors}}.
\newblock {\emph{\JournalTitle{{IEEE} Trans. Mag.}}}
  \textbf{\bibinfo{volume}{25}}, \bibinfo{pages}{2670--2677},
  \doiprefix\url{http://doi.org/10.1109/20.24507} (\bibinfo{year}{1989}).

\bibitem{Murgatroyd2000}
\bibinfo{author}{Murgatroyd, P.~N.} \& \bibinfo{author}{Eastaugh, D.~P.}
\newblock \bibinfo{journal}{\bibinfo{title}{Optimum shapes for multilayered
  toroidal inductors}}.
\newblock {\emph{\JournalTitle{IEE Proc. Electric Power Appl.}}}
  \textbf{\bibinfo{volume}{147}}, \bibinfo{pages}{75--81},
  \doiprefix\url{http://doi.org/10.1049/ip-epa:20000001}
  (\bibinfo{year}{2000}).

\bibitem{Polya1951}
\bibinfo{author}{P\'olya, G.} \& \bibinfo{author}{Szeg\H{o}, G.}
\newblock \emph{\bibinfo{title}{Isoperimetric inequalities in mathematical
  physics}}, vol.~\bibinfo{volume}{27} of \emph{\bibinfo{series}{Annals of
  Mathematics Studies}} (\bibinfo{publisher}{Princeton University Press},
  \bibinfo{address}{Princeton}, \bibinfo{year}{1951}).

\bibitem{Rayleigh1912}
\bibinfo{author}{Rayleigh, L.}
\newblock \bibinfo{journal}{\bibinfo{title}{On the self-induction of electric
  currents in a thin anchor-ring}}.
\newblock {\emph{\JournalTitle{Proc. Roy. Soc. A}}}
  \textbf{\bibinfo{volume}{86}}, \bibinfo{pages}{562--571},
  \doiprefix\url{http://doi.org/10.1098/rspa.1912.0046} (\bibinfo{year}{1912}).

\bibitem{Butterworth1925}
\bibinfo{author}{Butterworth, S.}
\newblock \bibinfo{journal}{\bibinfo{title}{On the alternating current
  resistance of solenoidal coils}}.
\newblock {\emph{\JournalTitle{Proc. Roy. Soc. A}}}
  \textbf{\bibinfo{volume}{107}}, \bibinfo{pages}{693--715},
  \doiprefix\url{http://doi.org/10.1098/rspa.1925.0050} (\bibinfo{year}{1925}).

\bibitem{Howe1917}
\bibinfo{author}{Howe, G. W.~O.}
\newblock \bibinfo{journal}{\bibinfo{title}{The {H}igh-{F}requency {R}esistance
  of {M}ultiply-{S}tranded {I}nsulated {W}ire}}.
\newblock {\emph{\JournalTitle{Proc. Roy. Soc. A}}}
  \textbf{\bibinfo{volume}{93}}, \bibinfo{pages}{468--492},
  \doiprefix\url{http://doi.org/10.1098/rspa.1917.0033} (\bibinfo{year}{1917}).

\bibitem{Rayleigh1886}
\bibinfo{author}{Rayleigh, L.}
\newblock \bibinfo{journal}{\bibinfo{title}{{LII}. {O}n the self-induction and
  resistance of straight conductors}}.
\newblock {\emph{\JournalTitle{The London, Edinburgh, and Dublin Philosophical
  Magazine and Journal of Science}}} \textbf{\bibinfo{volume}{21}},
  \bibinfo{pages}{381--394},
  \doiprefix\url{http://doi.org/10.1080/14786448608627863}
  (\bibinfo{year}{1886}).

\bibitem{Smith1972}
\bibinfo{author}{Smith, G.~S.}
\newblock \bibinfo{journal}{\bibinfo{title}{Proximity effect in systems of
  parallel conductors}}.
\newblock {\emph{\JournalTitle{{J}. {A}ppl. {P}hys.}}}
  \textbf{\bibinfo{volume}{43}}, \bibinfo{pages}{2196--2203},
  \doiprefix\url{http://doi.org/10.1063/1.1661474} (\bibinfo{year}{1972}).

\bibitem{Burghartz2003}
\bibinfo{author}{Burghartz, J.~N.} \& \bibinfo{author}{Rejaei, B.}
\newblock \bibinfo{journal}{\bibinfo{title}{On the design of {RF} spiral
  inductors on silicon}}.
\newblock {\emph{\JournalTitle{{IEEE} {T}rans. {E}lectron {D}ev.}}}
  \textbf{\bibinfo{volume}{50}}, \bibinfo{pages}{718--729},
  \doiprefix\url{http://doi.org/10.1109/ted.2003.810474}
  (\bibinfo{year}{2003}).

\bibitem{Tumanski2007}
\bibinfo{author}{Tumanski, S.}
\newblock \bibinfo{journal}{\bibinfo{title}{Induction coil
  sensors{\textemdash}a review}}.
\newblock {\emph{\JournalTitle{{M}eas. {S}ci. {T}ech.}}}
  \textbf{\bibinfo{volume}{18}}, \bibinfo{pages}{R31--R46},
  \doiprefix\url{http://doi.org/10.1088/0957-0233/18/3/r01}
  (\bibinfo{year}{2007}).

\bibitem{Karalis2008}
\bibinfo{author}{Karalis, A.}, \bibinfo{author}{Joannopoulos, J.~D.} \&
  \bibinfo{author}{Solja{\v{c}}i{\'{c}}, M.}
\newblock \bibinfo{journal}{\bibinfo{title}{Efficient wireless non-radiative
  mid-range energy transfer}}.
\newblock {\emph{\JournalTitle{{A}nn. {P}hys.}}}
  \textbf{\bibinfo{volume}{323}}, \bibinfo{pages}{34--48},
  \doiprefix\url{http://doi.org/10.1016/j.aop.2007.04.017}
  (\bibinfo{year}{2008}).

\bibitem{Bahl2001}
\bibinfo{author}{Bahl, I.}
\newblock \bibinfo{journal}{\bibinfo{title}{High-performance inductors}}.
\newblock {\emph{\JournalTitle{{IEEE} Transactions on Microwave Theory and
  Techniques}}} \textbf{\bibinfo{volume}{49}}, \bibinfo{pages}{654--664},
  \doiprefix\url{http://doi.org/10.1109/22.915439} (\bibinfo{year}{2001}).

\bibitem{Smith}
\bibinfo{author}{Smith, G.}
\newblock \bibinfo{title}{The proximity effect in systems of parallel
  conductors and electrically small multiturn loop antennas}.
\newblock \bibinfo{type}{Tech. Rep.} \bibinfo{number}{624},
  \bibinfo{institution}{Division of Engineering and Applied Physics, Harvard
  University} (\bibinfo{year}{1971}).
\newblock \doiprefix\url{https://apps.dtic.mil/dtic/tr/fulltext/u2/736984.pdf}.
\newblock \bibinfo{note}{(unpublished)}.

\bibitem{Landau1975}
\bibinfo{author}{Landau, L.~D.}
\newblock \emph{\bibinfo{title}{The classical theory of fields}}
  (\bibinfo{publisher}{Pergamon Press}, \bibinfo{address}{Oxford New York},
  \bibinfo{year}{1975}).

\bibitem{Grover1946}
\bibinfo{author}{Grover, F.~W.}
\newblock \emph{\bibinfo{title}{Inductance {C}alculations}}
  (\bibinfo{publisher}{Dover Publications, Inc.}, \bibinfo{year}{1946}).

\bibitem{MATLAB2018}
\bibinfo{author}{MATLAB}.
\newblock \emph{\bibinfo{title}{9.7.0.1190202 (R2019b)}}
  (\bibinfo{publisher}{The MathWorks Inc.}, \bibinfo{address}{Natick,
  Massachusetts}, \bibinfo{year}{2019}).

\bibitem{Simpsons2001}
\bibinfo{author}{Simpsons, J.}, \bibinfo{author}{Lane, J.},
  \bibinfo{author}{Immer, C.} \& \bibinfo{author}{Youngquist, R.}
\newblock \bibinfo{title}{Simple analytic expressions for the magnetic field of
  a circular current loop}.
\newblock \bibinfo{type}{Tech. Rep.} \bibinfo{number}{NASA/TM-2013-217919},
  \bibinfo{institution}{NASA} (\bibinfo{year}{2001}).
\newblock
  \doiprefix\url{https://ntrs.nasa.gov/archive/nasa/casi.ntrs.nasa.gov/20010038494.pdf}.

\end{thebibliography}
\end{document}